\documentstyle{mn}
\input psfig.tex
\def\refitem{\par\parskip 0pt\noindent\hangindent 20pt}

\def\cm{\rm cm} 
\def\cm3{\rm cm^{-3}} 
\def\ergs{\rm erg\, s^{-1}}

\def\spose#1{\hbox to 0pt{#1\hss}} 
\def\approxlt{\mathrel{\spose{\lower 3pt\hbox{$\sim$}}\raise 2.0pt\hbox{$<$}}}
\def\approxgt{\mathrel{\spose{\lower 3pt\hbox{$\sim$}}\raise 2.0pt\hbox{$>$}}}

\title[Thermal material in jets] {Thermal material in relativistic jets}

\author[A. Celotti, Z. Kuncic, M.J. Rees, J.F.C. Wardle]
{A. Celotti$^1$\thanks{E-mail: {\tt celotti@sissa.it}},
Z. Kuncic$^{2}$, M.J. Rees$^3$, J.F.C. Wardle$^4$ \\ 
$^1$S.I.S.S.A., via Beirut 2--4, 34014 Trieste, Italy \\
$^2$Australian National University Astrophysical Theory Centre, ACT 0200, 
Australia\thanks{The ANUATC is operated jointly by the Mount Stromlo and
Siding Springs Observatories and the School of Mathematical Sciences.} \\
$^3$Institute of Astronomy, Madingley Road, Cambridge CB3 0HA \\
$^4$Physics Department, Brandeis University, Waltham, MA 02254, USA\\}

\begin{document} 
\maketitle 

\begin{abstract} The properties of thermal material coexisting with
non--thermal emitting plasma and strong magnetic field in the powerful
jets of Active Galactic Nuclei (AGN) are examined.  Theoretical and
observational constraints on the physical properties of this `cold'
component are determined. While the presence of a thermal component
occupying a fraction $\sim 10^{-8}$ of the jet volume is possible, it
seems unlikely that such a component is capable of contributing
significantly to the total jet energy budget, since the thermal
reprocessing signatures that should appear in the spectra have not,
as yet, been detected.
\end{abstract}


\section{Introduction}

Powerful, highly--collimated jets with relativistic flow speeds transport
prodigious amounts of energy from the compact nuclei
of active galaxies. According to the standard paradigm (Blandford \& Rees
1978, Blandford \& K\"onigl 1979), these relativistic jets reveal
themselves chiefly through emission by a non--thermal distribution of
relativistic particles that are probably confined and also accelerated
{\em in situ} by strong magnetic fields. While little is known of any
other particle component which might contribute significantly to the
global energetics and internal structure of jets in AGN, it is possible
that there may be present, at least in small amounts, relatively cool
material with a quasi--thermal distribution. The question then arises: to
what extent can such material contribute to the reservoir of kinetic
energy that powers the relativistic jets? 

The answer to this question, apart from offering much needed insight into
the physical conditions in relativistic jets, may also have important
implications for the connection between jets in AGN and those in young
stars and Galactic compact objects. Although morphologically similar to
AGN jets, the jets in young stars are largely `thermal' (e.g. Herbig Haro
objects are seen mainly in emission lines rather than continuum emission)
and are moving with sub--relativistic flow speeds (see Wilson 1993 for a
review). Jets in Galactic compact objects generally exhibit properties that
are intermediate between those of jets in young stars and AGN, emitting both
thermal and non--thermal radiation; they move with flow speeds that are much
higher than in young stars, sometimes reaching the relativistic velocities
(e.g. the binary system SS~433, Vermeulen 1993; see also Mirabel \& Rodriguez
1996) that are characteristic of AGN. 

In AGN jets, it is plausible that at least some quasi--thermal material is
present simply because the acceleration of particles to relativistic
energies is unlikely to be 100 per cent efficient. Indeed, the
acceleration might not be continuous; non--thermal particles that have
cooled down (or initially constitute the low-energy end of the particle
distribution spectrum) can thermalize before they are re--accelerated
(e.g. Ghisellini, Guilbert \& Svensson 1988; de Kool, Begelman \& Sikora
1989). Some material might also be trapped at the base of jets as they
form, while some is expected to be entrained from the external medium as
jets propagate, following dynamical instabilities at the boundary. If this
cooler material accumulates to high enough densities, the thermalization
and radiative timescales can be much shorter than the dynamical
timescales. Such thermal gas should then be capable of producing spectral
signatures. 

The optical--UV spectral energy distributions of radio--loud quasars
(and those of their radio--quiet counterparts), suggest that there
is a significant amount of thermal material, in the form of broad and
narrow line emitting gas, at least in the immediate environments of jets
(Wills 1996). In cases where spectral signatures are seen, however,
this reprocessing gas is inferred to reside close to, but outside 
the jets (see Wilson 1993 for a review and also Walker, Romney \&
Benson 1994). This is most spectacularly shown by the HST images of
narrow line emitting gas (Capetti et al. 1996).

To date, there exists no radiative or dynamical evidence to suggest that
there may be thermal material actually inside relativistic jets in AGN. 
The most extreme case is with BL Lacs. Despite sharing common spectral
properties with radio--loud quasars (e.g. power--law continua, rapid
variability, high degrees of linear polarization -- see Bregman 1990),
their essentially featureless spectra offer few clues to the possibility
that thermal matter may be present in or near their jets.  It still
remains unclear whether this is due to the powerful jet emission swamping
any underlying reprocessing features or whether these objects are indeed
devoid of significant amounts of thermal emitting plasma or ionizing
radiation.  Recent observations of weak broad emission lines in several
sources (Stickel, Fried \& K\"uhr 1993;  Vermeulen et al. 1995;  Robinson
\& Corbett 1996;  Nesci \& Massaro 1996)  have provided some evidence for
thermal gas associated with BL Lacs. Also, the presence of relatively cold
outflowing material has been inferred from the observations of absorption
features in the extreme ultraviolet (EUV)  (K\"onigl et al. 1995) and soft
X--ray bands (Canizares \& Kruper 1984, Krolik et al. 1985, Madejski et
al.  1991)\footnote{Recent observations with the SAX satellite also show
signs of absorption in the soft X--ray band of the radio--loud quasar 3C
273, Grandi et al. 1997).}. However, at most mildly relativistic
velocities are derived, again suggesting that the absorbing gas is likely
located outside the jet or at its boundaries. 

It is possible that the thermal matter content of jets in radio--loud
quasars and BL Lacs is regulated by the structure of the internal
magnetic field.  Together with the non--thermal emitting particles and
the strong radiation field, the magnetic field is likely to be a
dominant component in the jet environment. Indeed, it is expected that
the internal structure of jets becomes complex and inhomogeneous on
small scales owing to its presence. The particle, radiation and
magnetic field components plausibly coexist in a multiphase
medium, similar to what is believed to prevail in the central
magnetosphere and in the more distant line--emitting regions of AGN. A
spectacular example of this complex internal structure comes from the
high resolution observations of the jet in M87 (Biretta et
al. 1995). These images reveal the presence of both large, regular
emission sites and weaker substructure resembling magnetic filaments.

The role of magnetic fields in confining cool `clouds' at the very centres
of AGN was first pointed out by Rees (1987) and subsequently considered by
Celotti, Fabian \& Rees (1992). More recently, the physical and radiative
properties of clouds of cool, magnetically confined material in a
typical AGN magnetosphere have been studied in detail by Kuncic, Blackman
\& Rees (1996) and Kuncic, Celotti \& Rees (1997). While some physical
effects are similar to the case studied here, the environment, the
dynamics, and the observational constraints are obviously quite different
in the jet case. 

In fact, in most or all of the observed wavebands, from radio to
$\gamma$--rays, the spectral energy distribution of AGN known to have
relativistic collimated jets seems to be dominated by radiation produced
in the jet itself. The most likely radiation mechanisms reproducing the
characteristics of the observed non--thermal spectra are synchrotron
emission and inverse Compton scattering by relativistic
electrons\footnote{Interestingly, it has been argued (Eilek \& Caroff
1979) that these cooling processes in particular can stabilize
relativistic plasma against thermal instability triggered by the presence
of localized regions of cool gas.}. The scattering can be off either the
synchrotron photons or a radiation field external to the jet, such as the
soft photons from an underlying accretion disk or line emission scattered
by diffuse electrons (see e.g. Sikora, Begelman \& Rees 1994).

In this paper, we determine limits on the amount and physical properties
of quasi--thermal cold material in a relativistic jet environment. We
consider both observational constraints, which so far only allow us to set
upper limits, and theoretical ones, which are based on thermal, radiative
and dynamical arguments. The expected radiative signatures of thermal
matter and their detectability are also discussed. In Sections 2 and 3, we
consider constraints deriving from energetic, dynamical and radiative
considerations on a `macroscopic' level, while in Section 4, we briefly
discuss the limits imposed by the `microscopic' effects of diffusion.
Observational considerations and results based on polarization measures
are examined in Section 5. All the limits are compared and discussed in
Section 6, and conclusions are drawn in Section 7. 

\section{Energetic considerations} 

The strongest limit on the total amount of any thermal material contained
within a relativistic jet is imposed by the total kinetic power. If the
global energetics of jets is dominated by three main components, namely
electromagnetic fields, non--thermal relativistic particles and thermal
gas all comoving with a bulk speed $\sim c$, then the total jet power is
given by \begin{equation} L_{\rm jet}\approxgt L_{\rm em}+ L_{\rm nt} +
L_{\rm t} \simeq \pi \phi^2 R^2 \Gamma^2 c ({U_{\rm B}+U_{\rm p}}) \, ,
\end{equation} where $L_{\rm em}, L_{\rm nt}, L_{\rm t}$ are the powers in
the above three components, respectively, $\phi$ is the opening angle of
the jet, $\Gamma$ its bulk Lorentz factor and $R$ is a measure of
distance along the jet axis. The inequality takes into account any power
dissipated e.g. in the form of radiation.

The total power can be re-expressed in terms of the energy densities in
the field and particles, $U_{\rm B}$ and $U_{\rm p}$, respectively. 
$U_{\rm p}$ then represents the contribution of both the non--thermal
(relativistic) and thermal (non--relativistic) components, and critically
depends on the composition of the plasma. We assume that most protons are
cold and that the relativistic electrons have a power--law distribution
with typical energy index $\approxgt 2$ (in the following we will assume
it to be $\sim 2$, corresponding to a photon energy spectral index
$\alpha\sim 0.5$) and with a minimum Lorentz factor $\gamma_{\rm e, min}
\approxlt 10^3$. Under these assumptions, the power is dominated by the
bulk kinetic energy of the proton component, so $U_{\rm p} \sim m_{\rm p}
c^2 (n_{\rm nt} + f_{\rm v} n_{\rm t})$. In the case of a plasma composed
mainly of electron--positron pairs, $U_{\rm p} \sim 2 m_{\rm e} c^2
(n_{\rm nt} \langle\gamma_{\rm e}\rangle + f_{\rm v} n_{\rm t} m_{\rm
p}/m_{\rm e})$ where $\langle\gamma_{\rm e}\rangle$ is the average
electron Lorentz factor and $f_{\rm v}$ is the volume filling factor of
the thermal material (the non--thermal plasma is assumed to fill the
volume of the jet). 

According to eq.~(1), the total jet power implies a maximum {\it average}
gas density \begin{equation} \langle n_{\rm max, jet}\rangle \sim (7\times
10^{9}) L_{\rm jet,46} \phi_{-1}^{-2} R_{14}^{-2} \Gamma_{10}^{-2}
\qquad \cm3 \, , \end{equation} where we have assumed
reference values of $L_{\rm jet} = 10^{46} L_{\rm jet,46}\, \ergs$,
$R= 10^{14} R_{14}\,$ cm, $\Gamma\sim \delta = 10 \Gamma_{10}$ (where
$\delta$ is the Doppler factor) and $\phi=0.1 \phi_{-1} \simeq
\Gamma_{10}^{-1}$.  This suggests the possibility that an
energetically significant gas component could then be present in a
relativistic jet in the form of very small and extremely dense clumps
or `clouds'. Typical total jet luminosities of the order of $10^{46}$ erg
s$^{-1}$ have been assumed on the basis both of the total power
which can be extracted from a $10^8 M_{\odot}$ black hole and 
the estimate of the jet kinetic power on pc and extended scales
(e.g. Rawlings \& Saunders 1991; Celotti \& Fabian 1993).

We also expect magnetic fields to play a significant role. A tangled or
quasi--isotropic field component is likely to be responsible for the
acceleration of particles to non--thermal energies. On the other hand,
a large scale, ordered field, possibly anchored in an accretion disk,
may be responsible for the collimation and acceleration of jets
(see e.g. the review by Blandford 1993). The strength of a comoving
magnetic field that carries a power $L_{\rm jet}$ as Poynting flux is
typically \begin{equation} B\simeq (2\times 10^4) L_{\rm jet,46}^{1/2}
\phi_{-1}^{-1} R_{14}^{-1} \Gamma_{10}^{-1}\qquad {\rm G}. \end{equation}

Any thermalized material that becomes trapped by such a strong field in a
relativistic jet is bound to be subjected to dynamical forces and
instabilities, which would produce clumped structures comoving with the
jet. These structures would quickly come into pressure balance with the
surroundings and would effectively be confined by the field stresses. The
condition of pressure equilibrium then imposes a maximum density, which
for gas with an internal plasma parameter $\beta\gg 1$, is
\begin{equation} n_{\rm max,press} \sim (3\times 10^{17}) B_4^{* 2}
T_5^{-1} R_{14}^{-2} \qquad {\rm cm^{-3}}, \end{equation} where a
reference value $T\simeq 10^{5} T_5$ K has been assumed for the
temperature of the gas (see Section 3.1) and $B_4= 10^{-4} B$ G. Here and
in the following we indicate quantities estimated at $R_{14}$ with an
asterisk.  Note that the constraint above applies to the effective density
of the material, independent of its degree of `clumpiness'. A comparison
with the limit imposed by the total jet power shows that the condition of
pressure balance gives a more stringent limit on the density in cold
material, $n_{\rm t}$ for filling factors $f_{\rm v} \approxlt 10^{-8}$. 
This would then corresponds to $n_{\rm max,jet} \sim (7\times 10^{17}) 
L_{\rm jet,46} \phi_{-1}^{-2} R_{14}^{-2} \Gamma_{10}^{-2} f_{\rm
v,-8}^{-1}$ cm$^{-3}$ where $f_{\rm v, -8}= 10^8 f_{\rm v}$. 

In the following sections, we indeed argue that cold thermal gas is most
likely to occupy only a fraction $f_{\rm v} \ll 1$ of the jet volume.  In
other words, the magnetic fields in relativistic jets are capable of
providing an effective confining mechanism for an energetically
significant amount of gas at the very high densities that are compatible
with small volume filling factors.  Furthermore, we show (in Section 3.1)
that these densities are sufficiently high to sustain cool temperatures
whilst such `clouds' are immersed in the intense non--thermal radiation
field of the jet. 

\section{Macroscopic properties} 

Hereafter, we assume the presence of dense clouds in an AGN jet environment
at a distance $R$ from the central massive object under the condition of
particle and field conservation (i.e. $n_{\rm nt} \propto R^{-2}$ and
$B\propto R^{-1}$). Although the limits discussed so far are independent
of the specific spatial distribution or individual geometry of the clouds,
we assume for simplicity that they are spherical with a radius $r$.
Realistically, however, the clouds are likely to be filamentary or tube like,
elongated along the field direction, so that $r$ represents a measure of
the pressure scaleheight or the characteristic dimension transverse to
the magnetic field. We consider constraints on the possible physical
conditions (namely the density and dimension) for the cold clouds to exist
and survive during their propagation along a jet. 

\subsection{Temperature and density}

Despite the intense radiation field which prevails in a relativistic jet
environment, any thermal material can rapidly equilibrate at the
equivalent blackbody temperature of the radiation field provided the gas
density is sufficiently high that two--body radiative processes become
more efficient than particle--photon interactions. For clouds comoving
with the jet, the photon field could be dominated either by the local
non--thermal radiation emitted by the ambient relativistic plasma or by
the quasi--thermal radiation external to the jet. 

An estimate of the cloud electron temperature can therefore be obtained by
considering the energy density of radiation as measured in the cloud
frame. This would be of the order of $U_{\rm rad} \sim (3\times 10^4) 
L_{\rm obs,46} \delta_{10}^{-3} \phi_{-1}^{-1} R_{14}^{-2}$ erg $\cm3$ (if
$R$ is estimated by variability timescales then $U_{\rm rad} \propto
\delta^{-5}$), where $L_{\rm obs,46}$ is a reference value for the
observed (beamed) radiative luminosity. On the other hand, an external
field of, say, $L_{\rm ext,45} = 10^{-45} L_{\rm ext}$ erg s$^{-1}$ in a
region of 10$^{15}$ cm, would lead to $U_{\rm rad, ext}\sim (3\times 10^5) 
L_{\rm ext,45} R^{-2}_{15} \Gamma_{10}^2$ erg $\cm3$. The equivalent
blackbody temperature would then be expected to be in the range $T_{\rm
bb} \sim (10^4-10^5)$ K. Clearly, the equilibrium temperature could be a
function of $R$, depending on the local radiation field. An attempt to
calculate a plausible radiative equilibrium temperature at various
distances by using the code {\small CLOUDY}, has lead to electron
temperatures in the range estimated above.  In the following a reference
value $T_5=10^{-5} T$ K will be used, at all $R$. 

This temperature can therefore be attained by clouds with densities such
that the bremsstrahlung cooling timescale is shorter than the Compton
heating timescale. The minimum density required for clouds to cool is thus
\begin{equation} n_{\rm min, cool} \simeq (2\times 10^{14}) U_{\rm rad}
T_{5}^{-1/2} \qquad \cm3 \, , \end{equation} where it has been assumed
that the local non--thermal field is energetically dominated by
$\gamma$--ray photons with energies $\sim m_{\rm e} c^2$. As blazar spectra
are typically dominated either by the synchrotron/blue bump component in
the optical--UV band or the Compton component at typically GeV energies,
the internal radiation field could in fact have a significantly different
Compton temperature and consequently the above estimate of $n_{\rm
min,cool}$ is uncertain. Furthermore the scattering radiation can be
mainly due to an external photon field (as in the case of high $\Gamma$).

Note that in the range of densities considered here, the radiative
timescales are smaller than the typical dynamical timescales, so that
spectral signatures could arise.

\subsection{Scaleheights} 

For pressure equilibrium to be maintained and restored after any
perturbation, clouds have to be sufficiently small that their
sound--crossing response time is shorter than the characteristic dynamical
timescale. This implies a maximum cloud size \begin{equation} r_{\rm max,
sound} \simeq 10^{9} R_{14} T_5^{1/2} \Gamma_{10}^{-1} \qquad {\rm cm} \,
.  \end{equation} However, any amount of material trapped in the jet
environment close to the central source would be subjected to strong
radiative, magnetic and inertial forces, the balance of which would
determine a typical scaleheight in the resultant direction. A cloud of
entrained material would feel the dynamic pressure from the bulk flow of
the jet which would be $\sim \Gamma^2 (c/v_{\rm s})^2 (n_{\rm nt}/n_{\rm
t}) g$, where $g$ is the gravitational acceleration and $v_{\rm s}$ the
sound speed, and would expand (in the comoving frame)  as this ram
pressure decreases, until the bulk speed is achieved. 

Forces due to radiation can also be dynamically important at the centres
of AGN. The intense radiation field in the innermost regions can prevent
the jet flow from rapidly accelerating to values of $\Gamma$ exceeding
about 10, due to the Compton drag felt by the non--thermal particles as
they scatter off aberrated photons, emitted from the inner regions of an
accretion disk or the more distant broad line region (Phinney 1987; see
also Sikora et al. 1996b). Indeed, most observational properties of AGN
jets (e.g. one--sidedness, superluminal proper motions and $\gamma$--ray
emission) can be accounted for with $\Gamma \sim 3 - 10$.  Radiation drag
might also be expected to affect any cooler, denser material immersed in
the intense radiation field. Such material, however, would be more
susceptible to radiative (bremsstrahlung) absorption than to Compton
scattering. If the low--frequency photons that are absorbed are mainly
those emitted by the ambient non--thermal plasma in the jet, the analogous
drag effect on the absorbing thermal material would be negligible (since
the synchrotron photons are highly anisotropic and therefore suffer little
relativistic aberration).  Instead, the material would feel a net
radiative acceleration $a\sim g \int {\rm d}\nu (\sigma_{\rm
abs,\nu}/\sigma_{\rm {\scriptscriptstyle T}})  (L_{\nu}/L_{\rm Edd})$
until it reaches its terminal velocity, where $\sigma_{\rm abs}$ and
$\sigma_{\rm {\scriptscriptstyle T}}$ are the absorption and scattering
cross sections, respectively, and $L_{\rm Edd}$ is the reference Eddington
luminosity. This acceleration depends on the radiation field but could
reach $10^{4-6} g$ (Kuncic et al. 1996; see also Ghisellini et al. 1990
for the analogous effect of synchrotron absorption). 

The detailed description of the dynamics of a cloud is quite complex and
beyond the scope of this paper. What is relevant here is that the
resultant acceleration $a$ can easily be several orders of magnitude
higher than $g$, leading to an estimate of the typical scaleheight
\begin{equation} h \sim {g\over a} h_{\rm grav} \simeq (6\times 10^{6})
{g\over a} M_8^{-1} T_5 R_{14}^2 \qquad {\rm cm} \, , \end{equation} where
$M= 10^8 M_{\odot} M_8$ is the mass of the central black hole.  This
implies that in the inner part of jets, thermal plasma can have
inhomogeneities on scales no larger than centimeters.

\subsection{Spatial configuration} 

As already mentioned, the spatial distribution of any thermal material in
a relativistic jet strongly depends on the structure of the magnetic
field. If the thermal gas is diamagnetic $(\beta \gg 1)$, it would not
necessarily follow the kinematics of the magnetic and relativistic
particle components. Such material, however, would be slowly penetrated by
the field; in this condition the dynamical coupling would be effective and
the compression of material would be limited by the internal field
amplification. 

Magnetic forces associated with poloidal fields could squeeze the plasma
into filaments with a small covering factor, which could then be
accelerated outwards along the field lines (e.g. Emmering et al. 1992). At
large distances, however, the transverse toroidal component of the field
dominates and this is favorable to cloud confinement by the tangential
stresses associated with the curvature of the field lines. We would then
expect a much larger covering factor over the jet cross--sectional area. 

Almost independently of the specific geometry, we can derive limits on the
distribution of thermal gas. By combining the constraint on the total
power derived in eq.~(2), which imposes a maximum particle density, with
the minimum density required by the condition that the plasma is able to
effectively radiate extra heat due to conduction from the relativistic
phase (see Section 4.1), we can estimate an upper limit on the volume
filling factor:  \begin{equation} f_{\rm v} \approxlt (7\times 10^{-6}) 
n_{\rm nt, 9}^{* -1} \langle\gamma_{\rm e}\rangle ^{-1} T_5^{1/2} L_{\rm
jet,46} \phi_{-1}^{-2} \Gamma_{10}^{-2} \, .\end{equation}  Here,
$\langle\gamma_{\rm e}\rangle m_e c^2$ is the average energy of the
relativistic particles (the Coulomb interactions are dominated by
non--thermal particles with lowest energies) and $n^*_{\rm nt} =10^9
n^*_{\rm nt,9}$ is their number density (at $R_{14}$).  Hereafter the
non--thermal particle density is estimated as the maximum allowed by the
limits imposed on the total jet power, assuming particle conservation (and
a constant acceleration efficiency at all $R$), i.e.  $n_{\rm
nt}(R)=10^{9} n^*_{\rm nt,9} R_{14}^{-2}$. 

Similar limits on $f_{\rm v}$ are also derived from the minimum density
required for clouds to cool in the strong radiation field, as given by
eq.~(5). Thus, any cool (and necessarily dense) gas could readily fill a
small fraction of the jet volume, the filling factor being roughly
comparable with the typical values for the reprocessing material
responsible for the broad line emission. 

\subsection{Dynamics} 

Cold material confined by a tangential field would tend to comove with the
bulk flow of a jet. However, in a situation where the cold component is in
relative motion, with speed $v_{\rm c}$ with respect to the relativistic
bulk flow, as during an acceleration phase, Kelvin--Helmholtz
instabilities can lead to an effective mixing of the cold and relativistic
phases and the consequent rapid heating of the cold gas \footnote{Whittle
\& Saslaw (1986) (and references therein) pointed out the stabilizing
effect of magnetic fields with respect to Kelvin--Helmholtz instability.}. 
The disruptive effect of the instability could be suppressed for densities
at which its typical growing time is longer than the cloud sound crossing
time, i.e. $n_{\rm min, K-H} \approxgt (5 \times 10^{16}) n^*_{\rm nt,9}
T_5^{-1} R^{-2}_{14} (v_{\rm c}/c)^2\, \cm3$.

Clouds could also be subjected to instability due to the curvature of the
magnetic field lines (analogous to the Rayleigh--Taylor instability for
ordinary fluids). However, this depends on the details of the local field
structure. Finally, shocks could form once the material reaches the highly
relativistic jet speed.

\section{Microphysical constraints} 

The survival of cool clouds in a relativistic jet can be also severely
threatened at a microphysical level by diffusion, which acts across the
large gradients in particle density, temperature and magnetic field at the
boundary of the confined gas. Note that, due to the strong magnetic field,
any thermal gas at densities $n_{\rm t}\approxlt 10^{18} \cm3$ can be
effectively treated as a globally neutral plasma confined by the field for
any $r$ larger than $\sim$ cm.

In this section we consider the effect of transfer of energy into the
clouds due to both the relativistic plasma surrounding the thermal gas and
the high energy radiation. Furthermore, the diffusion of the magnetic
field and the expansion of the gas along its direction impose lower limits
respectively on the dimension and mass of material not `evaporating'
within a dynamical timescale. 

\subsection{Diffusion}

Relativistic particles can readily penetrate a cool, thermal phase
embedded within a jet and then effectively transfer energy, providing a
source of heat, through Coulomb collisions.  This diffusive process would
ultimately heat an entire cloud of cool gas. However, on the radiative and
dynamical timescales relevant here, the diffusion would be effective only
over dimensions $d(t_{\rm brem}) \simeq (5\times 10^{7}) T_5^{1/4} n_{\rm
t,15}^{* -1} \langle\gamma_e\rangle R_{14}^2$~cm and $d(t_{\rm dyn}) 
\simeq 10^{10} R_{14}^{3/2} n_{\rm t,15}^{* -1/2} \langle\gamma_e\rangle
\Gamma_{10}^{-1/2}$ cm, respectively. An appropriate Coulomb factor of
$\ln\Lambda\sim 10$ and a thermal particle density compatible with the
constraints presented above have been assumed\footnote{In the direction
perpendicular to the magnetic field lines, the diffusion effect is
negligible owing to suppression by a factor equal to the ratio of
collisional to gyro frequencies.}. Clouds can then maintain their cold
temperature if significantly larger than these diffusion lengths. 

Alternatively, smaller clouds can survive and not `evaporate' if dense
enough to efficiently radiate the energy input provided by the
relativistic particles. The minimum density required is thus
\begin{equation} n_{\rm min, diff} \simeq 10^{15} n^*_{\rm nt,9}
\langle\gamma_e\rangle^{-1/2} T_5^{-1/2} R_{14}^{-2} \qquad \cm3 \,. 
\end{equation} In other words, material with sufficiently high densities
is not dramatically affected by the presence of the ambient relativistic
plasma. The latter, however, would lose some of its internal energy as
low--energy particles thermalize and effectively condense into the cool
clouds. 

An even more relevant heating agent, which is not affected by the presence
of a strong field, has been discussed by Kuncic et al. (1996)  in the case
of magnetospheric clouds. This is the Coulomb heating provided by
electron--positron pairs formed in clouds as a consequence of $\gamma \! 
-\! \gamma$ interactions. However, the observed copious $\gamma$--ray
emission from radio--loud AGN indicates that jets are optically thin to
photon--photon pair production, in natural agreement with the evidence of
relativistic motion of the emitting plasma. Furthermore, suggestions that
the observed $\gamma$--ray emission is due to the superposition of spectra
from compact regions in an inhomogeneous jet (e.g. Blandford \& Levinson
1995), seem to be at odds with the lack of observed reprocessed radiation
at lower energies. Heating of cool material by electron--positron pairs in
AGN jets is therefore unlikely to be important. 
 
Finally, as a consequence of a non zero resistivity, any cold plasma would
tend to diffuse and smooth gradients over the scale size of the field. 
This scale has then to be typically larger than $r_{\rm min,diffB}
\approxgt 20\ T_9^{-3/4} \Gamma_{10}^{-1/2} R_{14}^{1/2}$ cm to avoid the
field penetrating the clouds within a dynamical timescale.

On the other hand, the particles can propagate along the direction of
field lines. This implies that material confined in a magnetic
filamentary/tube--like structure would progressively disperse, typically
at the sound speed. Ultimately the filament can become so thin that
diffusion across the field is effective and it completely disperses on a
timescale $\approxlt t_{\rm dyn}$. Therefore, in order for the cold phase
to survive long enough to propagate in the jet, a minimum amount of
material has to be initially clumped. This corresponds to a minimum mass
(at $R_{14}$) only of the order of \begin{equation} M\approxgt 10^{6}
R_{14}^2 T_5^{1/2} n^*_{\rm t, 15} \Gamma_{10}^{-1}\qquad {\rm g} \,
.\end{equation} Note that the limit on the mass flux imposed by the
constraint on the total jet power corresponds to $\Gamma^{-1}_{10}$ the
mass accreted at roughly the Eddington limit, $\dot M \approxlt 10^{24}
L_{\rm jet,46} \Gamma^{-1}_{10}$ g, i.e. $2\times 10^{-2} M_{\odot}\ {\rm
yr}^{-1}$. 

\section{Observational constraints} 

Significant amounts of thermal gas should produce observable spectral
signatures which can provide a diagnostic of jet composition. Depending
on $R$, we expect different observational constraints to prevail. In this
Section, we examine the possible radiative signatures of thermal material
in AGN jets, in terms of emission, absorption and Faraday rotation. 

\subsection{Column density limit} X--ray and EUV spectral observations of
radio--loud quasars and BL Lacs have so far failed to reveal convincing
evidence of excess absorption, by material in the jet, with respect to
that produced by the Galactic hydrogen column density. Spectral features
in the BL Lac object PKS 2155--304 (and a few others BL Lacs) at EUV
(K\"onigl et
al. 1995) and soft X--ray energies (around 0.5-0.6 keV; e.g. Canizares \&
Kruper 1984, Krolik et al. 1985, Madejski et al. 1991) are best
interpreted as variable absorption by gas moving at most at mildly
relativistic speeds, possibly in a wind rather than in the jet itself.
Recently, some authors (e.g. Elvis et al. 1994, Cappi et al. 1997 and
references therein) have claimed the presence of excess absorption in soft
X--rays that may be intrinsic to high redshift radio--loud quasars, with
an $N_{\rm H}$ in the range $\sim (1-50)\times 10^{21}$ cm$^{-2}$.
However, this material too is inferred to be outflowing at subrelativistic
speeds and it is possibly located at large scales (Elvis et al. 1996). 
The lack of absorption in the X--ray band strongly limits the presence of
neutral material along the line of sight of most sources to typical
$N_{\rm H}\sim n_{\rm t} \phi R f_{\rm v}\approxlt 10^{20-21}$ cm$^{-2}$. 

Note that the energetics constraint given by eq.~(2) leads to an upper
limit on the column density of $N_{\rm H} \approxlt (7\times 10^{22}) 
L_{\rm jet, 46} \phi_{-1}^{-1} R_{14}^{-1} \Gamma_{10}^{-2}$ cm$^{-2}$,
which is higher than that allowed from observations, unless the plasma is
highly ionized or most of the X--ray radiation is produced at distances $R
\gg R_{14}$.  \footnote{Here and in the following we consider optical
depths in the direction perpendicular to the jet axis because of the
relativistic aberration of the observed emission. In other words we assume
the factor for the aberration correction $\delta \sin\theta \sim 1$.}

\subsection{Free--free absorption} 

Dense, cold material along the line of sight in an AGN jet would absorb
both thermal and non--thermal radiation up to a characteristic frequency
(for $\nu \approxlt (2\times 10^{16}) T_5 \delta_{10}$ Hz) 
\begin{equation}\nu_{\rm brem}(R)\simeq (8\times 10^{13})  T_5^{-3/4}
\delta_{10} n^*_{\rm t,15} f_{\rm v,-8}^{1/2} \phi_{-1}^{1/2}
R_{14}^{-3/2} \; {\rm Hz} \, , \end{equation} which is adequate for a high
covering factor, for which an absorption feature can be detected in the
observed spectrum. Highly clumped material could therefore easily absorb
radiation up to the optical band in the inner regions of the jet. 

In particular the observable effects of bremsstrahlung
absorption strongly
depend both on the covering factor of the material and on the
structure of the non--thermal emitting region. If the synchrotron spectrum
is produced in a single region and the clouds covering factor is
$\approxlt 1$, the non--thermal radiation would be absorbed below
$\nu_{\rm brem}$ (as estimated with $f_{\rm v}\approxlt 1$)  only in
proportion to the covered area. For higher covering factors, on the other
hand, the spectrum would tend flatten toward $\nu^{-2}$ below the
absorption frequency. More realistically, however, the observed radiation
may be due to the superposition of spectra produced at different positions
along an inhomogeneous jet (e.g. Blandford \& K\"onigl 1979; see K\"onigl
1989 for a review of the different models proposed). In this case, any
observable effect would of course depend on whether or not the absorbing
thermal gas is co--spatial with the location where most of the radiation
at that frequency is emitted. The detailed predictions here depend on the
assumed gradients in the jet emissivity.  In the case where thermal clouds
(with the same physical properties)  occupy a large fraction of the jet
volume, the expected effect is a steep cutoff below $\nu_{\rm brem}$,
proportional to the spatial extension of the cold material in the jet. 

High brightness temperatures typical of the non--thermal emission in
blazars suggest that induced Compton scattering can also be a competing
process in depleting the low frequency photons. However, this would
dominate over free--free absorption only for clouds with sizes $r
\approxlt (2\times 10^{3}) \nu_{12}^{-0.5} T_5^{3/2} T_{\rm b,12} n^*_{\rm
nt,9} R^{-3}_{14} n_{\rm t,15}^{* -2}$ cm, where $T_{\rm b} =
10^{12}T_{\rm b,12}$ K is the brightness temperature of the radiation
field at a frequency $\nu = 10^{12} \nu_{12}$ Hz.

\subsection{Synchrotron self--absorption}

Clearly bremsstrahlung absorption would affect the emitted
spectrum at each $R$ if $\nu_{\rm brem}\approxgt \nu_{\rm syn}$, where
$\nu_{\rm syn}$ is the synchrotron self--absorption frequency. If the
non--thermal emission is produced in an inhomogeneous jet structure, with
different regions contributing at different observed bands, the local
non--thermal radiation will be self absorbed up to a characteristic
frequency (e.g. Blandford \& K\"onigl 1979)  \begin{equation}\nu_{\rm
syn}(R)\simeq (2\times 10^{14}) \phi_{-1}^{1/3} \delta_{10}^{1/3}
(\gamma_{\rm e,min} n_{\rm nt}^{*} B^{* 2})^{1/3} R_{14}^{-1} \quad {\rm
Hz} \, , \end{equation} where $n_{\rm nt}$ and $B$ have been normalized to
the typical values derived from VLBI observations. 

\subsection{Faraday Rotation} 

Another signature of the presence of thermal material is its effect on the
polarization of the synchrotron radiation.  Its plane of polarization
would be rotated through an angle $\chi = (2\times 10^{4}) n_{\rm t}
f_{\rm v} B f_{\rm B} R \phi \delta^{2} \nu^{-2}\,\, {\rm rad},$ where
$f_{\rm B}$ takes into account reversals in direction of the magnetic
field. 

The importance of this effect at different distances $R$ can be estimated
by determining the frequency at which the rotation angle of the
polarization vector is, say $\chi \sim 1$ rad: \begin{equation} \nu_{\rm
F}(R)\simeq 10^{15} \phi_{-1}^{1/2} \delta_{10} B_4^{* 1/2} n_{\rm
t,15}^{* 1/2} f_{\rm v,-8}^{1/2} R_{14}^{ -1} f_{\rm B}^{1/2}\quad {\rm
Hz}.  \end{equation}

Near the base of the jet, however, the putative dense clouds are possibly
diamagnetic, confined by the magnetic field rather than penetrated by it,
and also their covering factor may be small.  Thus, we do not expect to
observe Faraday rotation at optical wavelengths, unless the clouds are
disrupted and fill the jet very quickly. On the other hand, at centimeter
radio wavelengths (whose emission region is at $R \approxgt 10^{19}$ cm),
such clouds would almost certainly fill the jet, i.e.  $f_{\rm v}\sim 1$.
Then at $R=10^{19}$ cm and $\nu = 5$ GHz, $\chi = (9\times 10^{-7}) n_{\rm
t}^{*} B_{4}^{*} f_{\rm B} \phi_{-1} \delta_{10}^{2} \quad {\rm rad}.$

VLBI observations at 5 GHz have established that in most BL Lac objects
the magnetic field in the radio jet is aligned transverse to the local jet
direction (Cawthorne et al. 1991; Gabuzda et al. 1994). Hence we can set
$\chi \approxlt 1$ rad, leading to an upper limit on the average density
of thermal material at the base of the jet, \begin{equation} n_{\rm t}^{*}
f_{\rm v} \approxlt 10^{6} B_{4}^{*,-1} f_{\rm B}^{-1} \phi_{-1}^{-1}
\delta_{10}^{-2} \quad {\rm cm^{-3}}. \end{equation}

The kinetic power corresponding to the above average density is $L_{\rm
kin,t} \approxlt (2\times 10^{42}) (B_{4}^{*} f_{B} \phi_{-1})^{-1}$ erg
s$^{-1}$. Hence, if $B_{4}^{*} f_{\rm B} \phi_{-1} L_{\rm jet,46}
\approxgt 2\times 10^{-4}$, then $L_{\rm kin,t} < L_{\rm jet}$. This means
that most of the energy in the jet is carried by the relativistic
particles and magnetic field, and that the bulk kinetic energy of the
thermal matter (and low energy relativistic electrons) is insufficient to
form a significant energy reservoir to compensate for radiative and
expansion losses. A conservative estimate of the dissipation in form of
observed radiation in BL Lacs would in fact lead to comoving luminosities
$L_{\rm rad, Comp} \sim (8\times 10^{42}) L_{\rm rad, 45} \delta^{-3}_{5}$
erg s$^{-1}$ (where a typical observed luminosity $L_{\rm rad,45}$ has
been assumed) while the luminosity carried by the relativistic emitting
component and magnetic field at pc scales imply $L_{\rm B}+L_{\rm kin,nt}
\approxgt 10^{44}$ erg s$^{-1}$ (e.g. Celotti \& Fabian 1993). We note
that these latter estimates strictly refer to the case of BL Lac objects,
i.e. low power and weak--lined sources. 

At face value eq.~(14) gives a powerful constraint on the presence of
thermal material in radio emitting jets, but it depends strongly on the
factor $f_{\rm B}$, which describes the reduction in Faraday rotation due
to reversals in field direction. In the bright VLBI knots, the field has a
large-scale ordering imposed by shocks (Aller et al. 1985; Wardle et al.
1994) and there is no information about the fine grained structure. In the
inter-knot regions, it is plausible to describe the field structure as
nearly randomly oriented cells of uniform magnetic field. Then $f_{\rm B}
\sim N^{-\frac{1}{2}} \sim p/0.7$ where $N$ is the number of cells in a
resolution element and $p$ is the observed fractional polarization. In
cases where the inter--knot polarization has been measured, $p$ is
typically a few per cent implying $f_{\rm B} \sim$ a few $\times 10^{-2}$. 

This value is probably appropriate for BL Lac objects where both the
predominantly transverse magnetic field and the magnetic field strength
inferred for the radio emitting region are consistent with conservation of
flux ($B \propto 1/R$) and a field of $\sim 10^{4}$ G at the base of the
jet (at $\sim 10^{14}$ cm). However, it may be seriously overestimated for
quasars, where the magnetic field is often longitudinal and may be
dominated by sheared loops field ($f_{\rm B} \simeq 0$), as described by
Begelman, Blandford \& Rees (1984). Indeed it is possible that the
observed differences in the VLBI polarization properties of quasars and BL
Lac objects (e.g. Roberts \& Wardle 1990;  Gabuzda et al. 1992) is
associated with the presence or absence of significant quantities of
thermal material in/around the jets. 

Finally, it is interesting to note that the estimates of the relativistic
particle density on these observational scales give $n_{\rm nt}\sim
10^{3} \gamma_{\rm e,min}^{-1}$ cm$^{-3}$, which is consistent with the
maximum density derived from the total power constraints for $\gamma_{\rm
e,min}\approxgt 100$ (e.g. Celotti \& Fabian 1993). 

\subsection{Emission}
 
Any absorbed radiation would be reprocessed by the thermal gas. It would
then
be typically re-emitted, as continuum, lines and edges, at an energy
$\delta kT$, with an intensity amplified by relativistic beaming. We refer
to Kuncic et al. (1997) for a detailed study of the radiative properties
and spectra emitted by clouds embedded in a non--thermal radiation field.
Most of the radiation is expected to be optically thin continuum and lines
plus edges concentrated in the EUV band and highly broadened by the bulk
motion of the emitting gas. 

Here, for simplicity, we estimate the emission by a thermal plasma at a
temperature $\sim 10^5T_5$, in the case of optically thin emission. This
would be observed as (beamed) radiation up to a cut--off frequency
$\nu_{\rm cut}\simeq (2\times 10^{16}) T_5 \delta_{10}$ Hz. This
luminosity can become energetically significant and potentially observable
if $\nu_{\rm cut} L_{\rm \nu, brem}$ (erg s$^{-1}$) is comparable with the
observed spectral power, say $\simeq 10^{46} L_{\rm rad,46}$. The
corresponding density of thermal gas that is required to produce $L_{\rm
rad,46}$ is then \[ n_{\rm t} \approxgt (4\times 10^{17}) L_{\rm rad,46}
^{1/2} \delta^{-3/2}_{10} T^{1/4}_5 \phi^{-1}_{-1} f_{\rm v,-8}^{-1/2}
R_{14}^{-3/2} \] \begin{equation} \hfill \exp(0.48\nu_{16}/\delta_{10}
T_5)^{1/2} \qquad \cm3 , \;  \end{equation} where the exponential term is
of the order of unity. 

A further constraint on the amount of thermal material in relativistic
jets can be derived from the absence of any observed Compton scattered
component in the soft X--ray band of radio--loud sources. In fact, Sikora
et al. (1996a) pointed out that the presence of cold electrons in the jet
could be detected through the effect of Comptonization of any external
radiation field. In the case of optical/UV blue bump photons, this
scattered component would contribute to the soft X--ray emission. On the
other hand, the lack of any spectral signature in this band imposes a
limit on the amount of cold material comoving with the jet, which can be
translated into $n_{\rm t}\approxlt (2\times 10^{17}) \phi_{-1}^{-2}
L_{\rm soft X,46} L_{\rm UV, 45}^{-1} \Gamma_{10}^{-2} f_{\rm v,-8}^{-1}
R_{14}^{-1}$ cm$^{-3}$. This limit again implies a minimum level of
`clumpiness', with values of $f_{\rm v}$ similar to those already derived,
at least for distances $R$ where an intense external radiation field is
likely to be present (typically out to $R \approxlt 0.1$ pc). 

\section{Summary and results}

In this Section, the limits derived so far are considered globally. In
particular, we compare the relative importance of emission, absorption and
Faraday rotation due to thermal material at different frequencies and
distances along a relativistic jet. 

The constraints on cool clouds of thermal material immersed in an
AGN jet are plotted in Fig.~1 in a density versus size ($n_{\rm t}$ vs. $r$)
parameter space diagram. The limits are estimated in the inner jet, at a
distance $R=
10^{14}R_{14}$ cm. In Fig.~2 only the tightest physical
constraints 
are 
reported at different
distances along the jet, namely $R= 10^{14}R_{14}, \, 10^{16}R_{16}, \,
10^{18}R_{18}$ cm, where jets become spatially resolved by VLBI (from the
top to the bottom panel, respectively). A filling factor of $f_{\rm
v}=10^{-8}$ has been adopted at all $R$. Each line is labelled as
explained in the figure caption of Fig.~1 and throughout the previous
sections. For
comparison, one can locate in the same plane the physical properties
inferred for broad--line clouds, i.e.  $n\sim 10^{8-11}$ cm$^{-3}$, $r\sim
10^{12}$ cm at $R\sim 0.1$ pc, for filling factors $\sim 10$ times higher
(e.g. Netzer 1990).

\begin{figure*}
\psfig{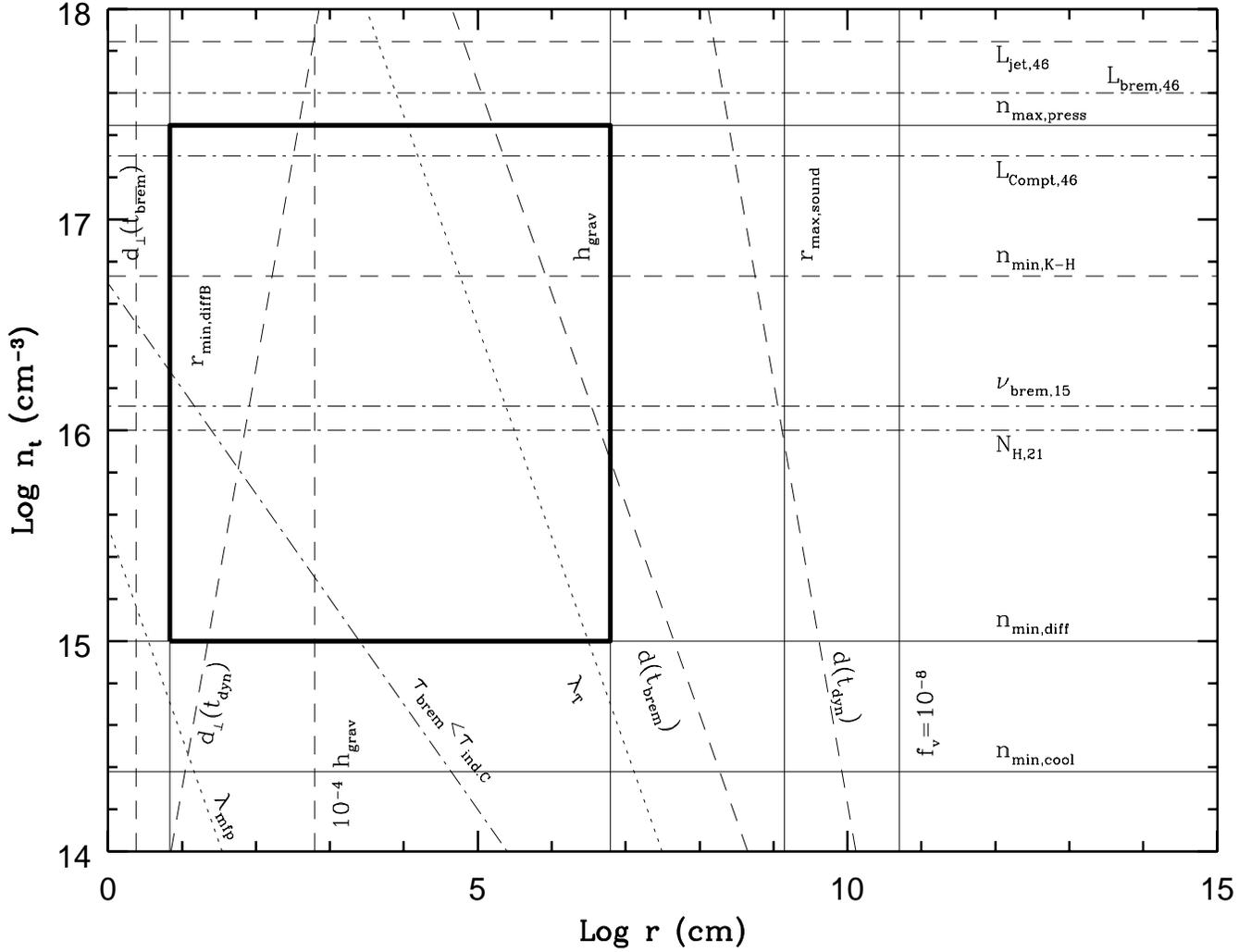}
\caption[h]
{Parameter space diagram (density, $n_{\rm t}$, versus thickness,
$r$) for thermal material in relativistic AGN jets at distance
$R = 10^{14} R_{14}$ cm,  with a temperature
$10^5$ K and a volume filling factor $10^{-8}$.  The various
constraints discussed at length in the text are plotted here and
identified by the following labels: \\
$n_{\rm max,press}$ --- density at which thermal pressure balances
confining magnetic pressure; \\
$n_{\rm min,diff}$ --- minimum density for gas to remain cool in the
presence of non--thermal diffusion; \\
$r_{\rm min,diffB}$ --- minimum scalesize of structure unaffected by
magnetic diffusion; \\
$h_{\rm grav}$ --- gravitational pressure scaleheight; \\
$r_{\rm max,sound}$ --- maximum sound--crossing scalesize for pressure
equilibrium on a dynamical timescale; \\
$L_{\rm jet,46}$ --- total jet power implies an upper limit on particle
kinetic energy density; \\
$L_{\rm brem,46}$ --- density responsible for bremsstrahlung
emission comparable to observed
spectral luminosity; \\
$L_{\rm Comp,46}$ --- density producing a Comptonization
luminosity comparable to
observed spectral power; \\
$n_{\rm min,cool}$ --- minimum density for 2--body processes to cool
gas more efficiently than particle--photon processes; \\
$n_{\rm min,K-H}$ --- minimum density at which Kevin--Helmholtz instability
is suppressed; \\
$N_{\rm H,21}$ --- neutral hydrogen column density limit;  \\
$\nu_{\rm brem,15}$ --- density corresponding to a spectral turnover at
optical frequencies due to 
bremsstrahlung absorption; \\
$d (t)$ --- non--thermal diffusion depth during a timescale $t$ ($t_{\rm dyn}$
and $t_{\rm brem}$ are the dynamical and cooling timescales);  \\
$d_{\perp} (t)$ --- same as $d (t)$, but for non--thermal diffusion transverse
to the magnetic field lines; \\
$\lambda_{\rm T}$ --- effective mean--free--path for collisions between
non--thermal and thermal particles;  \\
$\lambda_{\rm mfp}$ --- mean--free--path for collisions between thermal
particles only;  \\
$\tau_{\rm brem} < \tau_{\rm indC}$ --- relative optical depths due to
bremsstrahlung absorption and induced Compton scattering; \\
$f_{\rm v}$ --- (spherical) cloud dimension corresponding to a filling
factor
$f_{\rm v}$. 
Continuous lines indicate the strongest limits; dashed lines correspond
to constraints which can be overcome by different assumptions or cannot
be determined with precision; dash--dot lines show observational limits;
and dotted lines refer to physical quantities. The area enclosed by the
heavy set lines defines the resultant parameter subspace allowed by the
relevant constraints.}
\end{figure*}

\begin{figure*}
\psfig{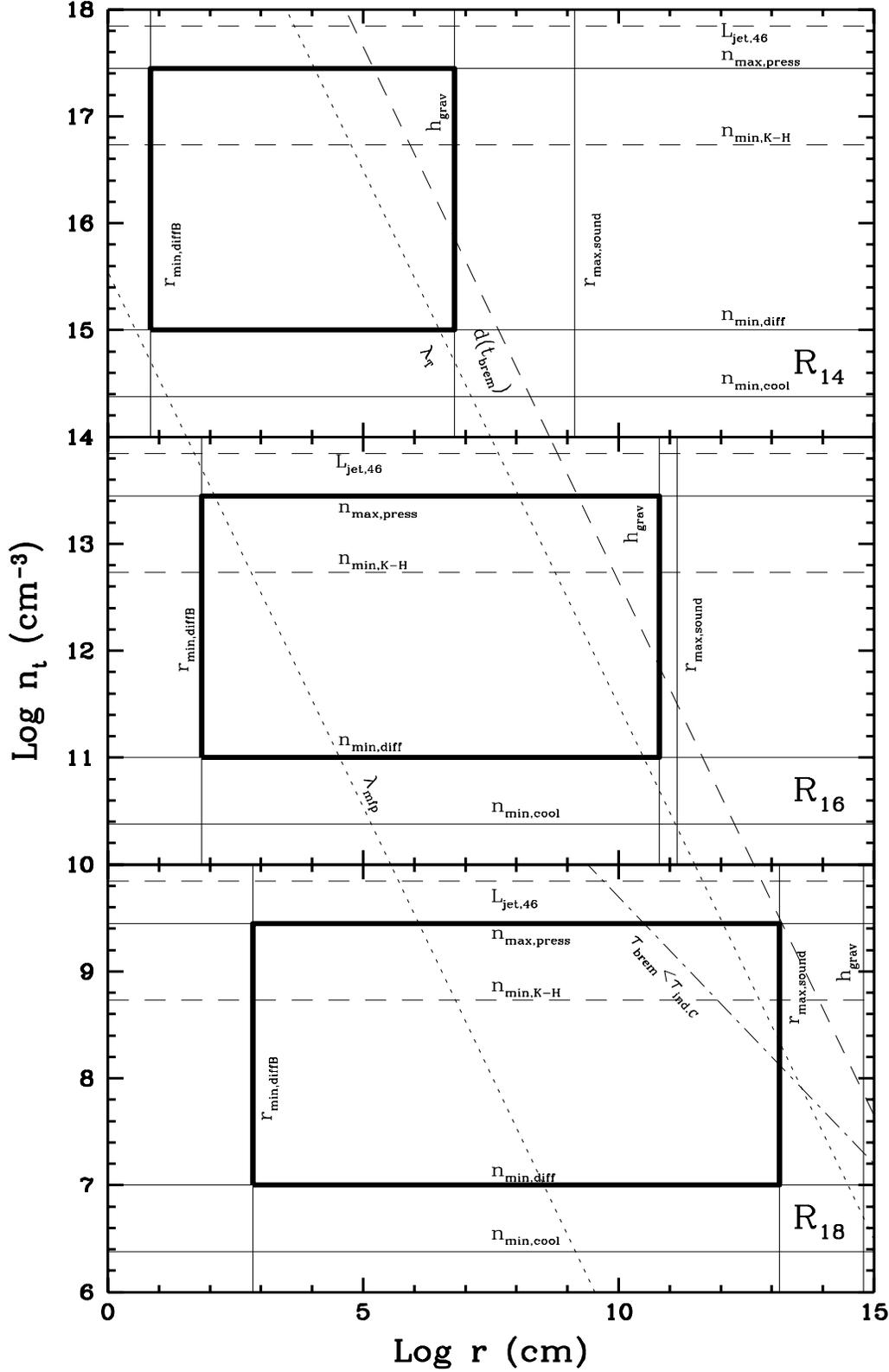}
\caption[h]
{Parameter space diagram (density, $n_{\rm t}$, versus thickness,
$r$) for thermal material in relativistic AGN jets at distances
$R = 10^{14} R_{14}$ cm (top panel), $10^{16} R_{16}$ cm (middle panel)
and $10^{18} R_{18}$ cm (bottom panel) and with a constant temperature
$10^5$ K and constant volume filling factor $10^{-8}$.  Only the 
tighter physical 
constraints are reported here. A complete
summary of the limits, including the observational ones, is included in
Fig.~1, with the same
labels and type of lines. The area enclosed by the
heavy set lines defines the resultant parameter subspaces allowed by the
relevant constraints.}
\end{figure*}

\subsection{Theoretical limits}

The gravitational scaleheight is the strongest upper limit on the cloud
size for $R\approxlt 10^{16}$ cm, with $h_{\rm grav} \sim 10^7$ cm (at
$R=10^{14}$ cm) and $\sim 10^{11}$ cm (at $R=10^{16}$ cm). As already
discussed, however, the typical scaleheight is likely to be affected by
much stronger radiative, dynamical and magnetic forces acting on the
clouds.  For effective accelerations of the order of $a\approxgt 10^{6-8}
g$, the existence of any clumped thermalized structure at these distances
is precluded. At $R>10^{16}$ cm, however, the requirement that the gas
responds to any perturbation on timescales smaller than $t_{\rm dyn}$
places a more stringent limit on $r$ than does $h_{\rm grav}$, giving
$r\approxlt 10^{13}$ cm. 

The lowest limits on $r$ are set by the magnetic field diffusion scales to
dimensions of the order of 10 to 10$^3$ cm. For low densities, the cloud
size starts to be comparable with the collisional mean free path and any
low frequency absorption is likely to be dominated by induced Compton
scattering rather than bremsstrahlung absorption. While thermal diffusion
from the relativistic phase can also act on such spatial scales, this
effect does not pose a serious threat to the survival of cool clouds
provided their density exceeds $n_{\rm min,diff}$ (which is estimated here
for a maximum density of the relativistic external plasma, $n_{\rm nt}
\sim 10^9$ cm$^{-3}$). Furthermore another constraint on the minimum
density of
thermal material is given by $n_{\rm min,cool}$, which is set by the
requirement that the gas can cool in the intense radiation field that is
expected to prevail within $10^{16-17}$ cm.

Finally, cloud pressures cannot exceed the confining magnetic one and this
constrains the thermal material to $n_{\rm t}\approxlt n_{\rm max,
press}$.  Note however that this has been calculated for diamagnetic
clouds and neglecting any internal radiation pressure.  It is also
relevant to stress here that both the pressure and the total jet power
limits are linearly dependent on the assumed jet power. 

The allowed parameter space is confined within the thick box in each panel
of Figs.~1 and 2.  The limits imply that the typical densities at which
cold
clumped matter can exist in the jet environment scale with distance as
$\sim R^{-2}$ and span about two--three orders of magnitude (at each
distance).

The other constraints discussed in the paper and reported in Figs.~1, 2
depend
on the volume filling factor as well as the dynamics of the clouds (e.g.
the Kelvin--Helmholtz stability limit). In general they indicate that
filling factors of the order of $10^{-8}$ are necessary for cold material
to have the range of densities required to survive in the jet environment. 

To summarize our findings, relatively cold, thermal material with small
volume filling factors can exist in the relativistic environment of an AGN
jet under the considered conditions at all scales $R\approxlt 1$ pc, when
confined and insulated by a strong magnetic field. Dynamically, these
structures can become unstable if they are subjected to an acceleration
phase on dynamical timescales. 

\subsection{Observational predictions}

Thermal material with the properties inferred above should produce
reprocessing signatures in the observed spectra.  In the inner jet, cool
material could absorb the locally produced radiation field up to optical
frequencies and re-emit this energy as quasi--thermal UV radiation at
levels comparable to the observed flux, for $f_{\rm v}\approxgt 10^{-8}$. 
This radiation would probably contribute and possibly exceed any component
from the optical to the EUV--soft X--ray spectral band in radio--loud
sources. 

At higher energies, thermal gas can manifest itself through photoelectric
absorption in the soft X--ray band. The observational limits on the column
density in neutral hydrogen are quite tight, typically $N_{\rm H}
\approxlt 10^{21}$ cm$^{-2}$, and are compatible with the range of gas
parameters found only if either a significant fraction of the cold
material is ionized or the covering factor is much lower than 1 or the
soft X--ray emitting region is not located in the inner part of the jet.
Furthermore, a Comptonized component is expected to be observable in the
soft X--ray band, from material with $f_{\rm v}\approxgt 10^{-8}$.  For
$R\approxgt 10^{16}$ cm, however, these radiative features are not
observable unless $f_{\rm v}\gg 10^{-8}$. 

Finally, we note that the observational constraints on Faraday rotation,
which are particularly relevant to BL Lac objects, correspond to the lower
end of the density range in each panel in Fig.~2, assuming that a filling
factor
$f_{\rm v}\sim 10^{-8}$ and $f_{\rm B}\sim 1$ remain constant at all $R$.
Values of $f_{\rm v}$ which are $\approxlt 10^{-9}$ or smaller values of
$f_{\rm B}$ would therefore be incompatible at all $R$ with the observational
results derived on scales $\sim 10^{19}$ cm (see however the discussion in
Section 5.4). We note that even if cooled electrons are not confined and
fill the jet, Faraday rotation measures could detect their presence.

\begin{figure} 
\psfig{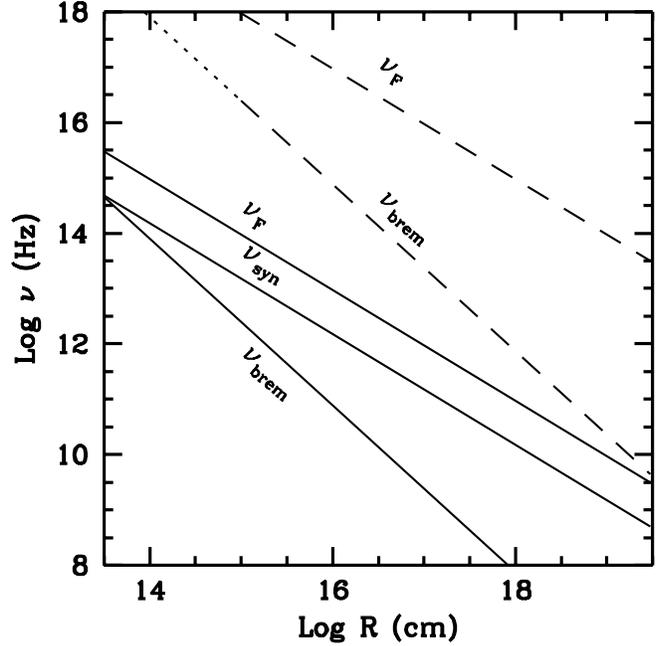}
\caption[h]{The dependence of the (observed) synchrotron ($\nu_{\rm syn}$) 
and bremsstrahlung ($\nu_{\rm brem}$) absorption frequencies and the
typical frequency at which Faraday rotation is measurable ($\nu_{\rm F}$)
on the distance $R$ along the jet. Volume filling factors $f_{\rm v}$,
possibly ranging from small values in the inner part of the jet ($\sim
10^{-8}$) to even $\sim 1$ at the observed scales, have been considered
and represented by continuous and dashed lines respectively. A Doppler
factor $\delta_{10}$ has been adopted. The dotted line section indicates
the bremsstrahlung absorption frequency computed in the Rayleigh--Jeans
limit (which overestimates $\nu_{\rm brem}$ at most by a factor $\sim 4$
at $R_{14}$)}
\end{figure}

Fig.~3 shows the relative importance of radiative absorption processes
and Faraday rotation at different frequencies and distances along the
jet, showing which radiative signature is most likely to be observable
in different spectral bands and at different resolution.

For densities of thermal material in the allowed parameters space of
Fig.~2, synchrotron self--absorption by the non--thermal relativistic
plasma is the most effective absorption process, for filling factors
of the thermal gas $f_{\rm v} \approxlt 10^{-8}$ (and the minimum thermal
gas density allowed by the limits reported in Fig.~2). Faraday rotation
can possibly affect radiation at frequencies a factor $\sim 4$ above the
synchrotron self--absorption ones. For higher values of $f_{\rm v}$,
as might be in the more external jet regions, Faraday rotation can be 
the strongest observational signature of the presence of thermal gas
(note however that the estimate of this effect is based on an assumed
$f_{\rm B}\sim 1$). Absorption by bremsstrahlung would be more effective
in depleting photons in the IR to UV bands up to a distance which depends
on $f_{\rm v}$.

As already discussed, any radiative signature (and in particular any
absorption feature) critically depends on the relative location of the
region mostly emitting at each frequency. In fact if the jet structure is
inhomogeneous, the observed spectra could be the superposition of the
emission from different regions of the jet, which contribute differently
at each frequency. As an example, in the parametric model developed by
Ghisellini \& Maraschi (1989) the region typically dominating the emission
in the near IR--optical--UV bands is located at about $R \sim 10^{16-17}$
cm, as suggested by simple variability timescale arguments (which in fact
sets upper limits on the size of the emitting region but not on its
location along the collimated jet). In this case then clouds would need to
have rather high covering and filling factors to produce detectable
signatures. 

\subsection{Energetic implications}

At all distances, the theoretical limits on the density of thermal
material are consistent with its kinetic power not exceeding the total jet
power, and imply that the gas cannot fill the jet for more than $\sim
10^{-5}$ of its volume.  For the adopted value of filling factor, its
contribution can be up to 25 per cent of the total power carried by the
jet.

From an observational point of view, however, the lack of thermal
signatures (in absorption or emission) from radio to soft X--ray
energies in radio--loud AGN, suggests that any thermal material is
confined in an even smaller volume.  Therefore the reservoir of energy
which can be carried as kinetic power of cold thermal gas is unlikely to
be energetically important, unless either $\Gamma\ll 10$ (which would make
AGN jets more closely resemble those of young stars or compact objects) or
some caveats about the lack of observed features are invoked.

\section{Conclusions}

There is strong evidence for the presence of at least 10--100 $M_{\odot}$
of thermal gas in the central regions of radio--loud AGN.  This
non--relativistic material may be closely associated with relativistic
jets either externally, as is probably the case with narrow line
emitting gas, or possibly internally, as is the case with jets
on stellar scales (e.g. SS~433).

Thermal material trapped in the central environs of AGN is likely to be
magnetically confined, as suggested by Rees (1987) and Emmering et al.
(1992) (see also the models by Ghisellini \& Madau 1996 for the
$\gamma$--ray emission and de Kool \& Begelman 1995 for broad absorption
line gas in quasars).  A strong magnetic field permits the presence of
gas that is multiphase in nature, with `cold' gas confined and thermally
insulated from non--thermal emitting plasma.  The spatial distribution of
the cold component is determined by the configuration of the field, so
that filamentary structures would probably prevail. Moreover, the boundary
of any material confined in such a way could be a preferential location for
acceleration of particles to relativistic energies, through e.g. magnetic
reconnection. 

In this paper, we have examined the physical conditions in the environment
of relativistic AGN jets containing both non--thermal and thermal matter
and we have found that the `cold' component could conceivably be present
at all distances throughout the jets, from the innermost regions out to
parsec scales, over several orders of magnitude in density and dimension.
The strongest constraints on the physical properties of such cold gas are
imposed by non--thermal diffusion effects associated with Coulomb collisions
with fast particles, the requirement of pressure equilibrium with the ambient
magnetic field, the gravitational pressure scaleheight and, on the smallest
scales, magnetic diffusion effects.
Unfortunately, the lack (so far) of radiative spectral signatures, such as
emission and absorption features as well as Faraday rotation measures, in the
observed non--thermal spectra limits the filling factor of thermal gas to
such small values that it cannot constitute an important component in the
overall energy budget of jets in radio--loud AGN.  We hope that future
broadband, spectroscopic and polarimetric observations of blazars will either
further constrain the above result or possibly reveal the predicted thermal
reprocessing features. 

We note that while in our analysis we have adopted a reference jet power
of $10^{46}$ erg s$^{-1}$, the constraints would be even stronger if higher
luminosities (as estimated in some powerful jets) were considered. It is
also important to point out that our findings are especially relevant for
thermal material in the jet of weak--lined sources as BL Lacs. In fact,
both their estimated total power is lower and observational constraints
(in terms of Faraday limits and optical/UV thermal emission) are certainly
tighter. This seems to support the view that indeed the environment of
weak blazars lacks significant amounts of thermal material. 

The absence of an energetically significant cold thermal component is an
important constraint on jet models, somehow in agreement with several
other suggestions of the deficiency of low energy particles. All these
indications point to a highly efficient (re--)acceleration mechanism
operating in the jet, either in form of e.g. reconnection, shocks or
heating by radiative absorption, which would have to maintain the bulk of
particles at least at mildly relativistic energies.

\section*{Acknowledgments} 
We acknowledge Gary Ferland for the use of his code {\small CLOUDY}.
For financial support, thanks are due to the Italian MURST (AC), the
Australian DIST (ZK), the Royal Society (MJR). JFCW acknowledges the
John Simon Guggenheim Memorial Foundation for a Fellowship.

\section*{References}

\refitem Aller H.D., Aller M.F., Hughes P.A., 1985, ApJ, 298, 296

\refitem Begelman M.C., Blandford R.D., Rees M.J., 1984, Rev. Mod. Phys.,  
56, 255

\refitem Biretta J.A., Sparks W.B., Macchetto F., Capetti A., 1995, BAAS,
187, 5016

\refitem Blandford R.D., 1993, in Burgarella D., Livio M., O'Dea C.P.,
eds, Astrophysical Jets. Cambridge Univ. Press, Cambridge, p. 15

\refitem Blandford R.D., K\"onigl A., 1979, ApJ, 232, 34

\refitem Blandford R.D., Levinson A., 1995, ApJ, 441, 79

\refitem Blandford R.D., Rees M.J., 1978, in Wolfe A.M., ed,
Proc. Pittsburgh Conf. on BL Lac Objects. Univ. of Pittsburgh Press,
Pittsburgh, p. 328

\refitem Bregman J.N., 1990, A\&AR, 2, 125

\refitem Canizares C.R., Kruper J., 1984, ApJ, 278, L99

\refitem Capetti A., Axon D.J., Macchetto F., Sparks W.B., Boksenberg
A., 1996, ApJ, 469, 554

\refitem Cappi M., Matsuoka M., Comastri A., Brinkmann W., Elvis M.
Palumbo G.G.C., Vignali C., 1997, ApJ, in press

\refitem Cawthorne T.V., Wardle J.F.C., Roberts D.H., Gabuzda D.C., 1993,
ApJ, 416, 519 

\refitem Celotti A., Fabian A.C., 1993, MNRAS, 264, 228

\refitem Celotti A., Fabian A.C., Rees M.J., 1992, MNRAS, 255, 419

\refitem de Kool M., Begelman M.C., 1995, ApJ, 455, 448

\refitem de Kool M., Begelman M.C., Sikora M., 1989, ApJ, 337, 66

\refitem Eilek J.A., Caroff L.J., 1979, ApJ, 233, 463

\refitem Elvis M., Fiore F., Wilkes B.J., McDowell J.C., Bechtold J., 
1994, ApJ, 425, 103

\refitem Elvis M., Mathur S., Wilkes B.J., Fiore F., Giommi P., Padovani
P., 1997, in Proc. of IAU Coll. 159, Emission Lines in Active Galaxies:
New Methods and Techniques, in press

\refitem Emmering R.T., Blandford R.D., Shlosman I., 1992, ApJ, 385, 460

\refitem Gabuzda D.C., Cawthorne T.V., Roberts D.H., Wardle J.F.C.,
1992, ApJ, 338, 40

\refitem Gabuzda D.C., Mullan C.M., Cawthorne T.V., Wardle J.F.C., 
Roberts D.H., 1994, ApJ, 435, 140

\refitem Ghisellini G., Madau P., 1996, MNRAS, 280, 67

\refitem Ghisellini G., Maraschi L., 1989, ApJ, 340, 181

\refitem Ghisellini G., Guilbert P.W., Svensson R., 1988, ApJ, 334, L5

\refitem Ghisellini G., Bodo G., Trussoni E., Rees M.J., 1990, ApJ, 362, L1

\refitem Grandi P., et al., 1997, A\&A, submitted 

\refitem K\"onigl A., 1989, in Maraschi L., Maccacaro T., Ulrich M.--H.,
eds, BL Lac Objects. Springer--Verlag, p.~321

\refitem K\"onigl A., Kartje J.F., Kahn S.M., Chorng--Yuan Hwang, 1995,
ApJ, 446, 598

\refitem Krolik J.H., Kallman T.R., Fabian A.C., Rees M.J., 1985,
ApJ, 295, 104

\refitem Kuncic Z., Blackman E., Rees M.J., 1996, MNRAS, 283, 1322

\refitem Kuncic Z., Celotti A., Rees M.J., 1997, MNRAS, 284, 717

\refitem Madejski G.M., Mushotzky R.F., Weaver K.A., Arnaud K.A., Urry 
C.M., 1991, ApJ, 370, 198

\refitem Mathews W.G., Ferland G.J., 1987, ApJ, 323, 456

\refitem Mirabel I.F., Rodriguez L.F., 1996, in Tsinganos K.C., ed, 
Solar and Astrophysical Magnetohydrodynamical Flows. Kluwer Academic 
Publisher, The Netherlands, p. 683

\refitem Nesci R., Massaro E., 1996, IAU Circ. 6457

\refitem Netzer H., 1990, in Courvoisier T.J.-L., Mayor M., eds, Active
Galactic Nuclei, 20$^{th}$ SAAS--FEE Course. Springer--Verlag. 

\refitem Phinney E.S., 1987, in Zensus J.A., Pearson T.J., eds,
Superluminal Radio Sources. Cambridge Univ. Press, Cambridge, p. 301

\refitem Rawlings S.G., Saunders R.D.E., 1991, Nat, 349, 138

\refitem Rees M.J., 1987, MNRAS, 228, 47p

\refitem Roberts D.H., Wardle J.F.C., 1990, in Zensus J.A., Pearson T.J., eds,
Superluminal Radio Sources. Cambridge Univ. Press, Cambridge, p. 193

\refitem Robinson A., Corbett E., 1996, IAU Circ. 6463

\refitem Sikora M., Begelman M.C., Rees M.J., 1994, ApJ, 421, 153

\refitem Sikora M., Madejski G., Moderski R., Poutanen J., 1996a, ApJ, 
submitted

\refitem Sikora M., Sol H., Begelman M.C., Madejski G.M., 1996b, MNRAS,
280, 781

\refitem Stickel M., Fried J.W., K\"uhr H., 1993, A\&AS, 98, 393

\refitem Vermeulen R.C., 1993, in Burgarella D., Livio M., O'Dea C.P.,
eds, Astrophysical Jets. Cambridge Univ. Press, Cambridge, p. 241

\refitem Vermeulen R.C., Ogle P.M., Tran H.D., Browne I.W.A.,
Cohen M.H., Readhead A.C.S., Taylor G.B., 1995, ApJ, 452, L5 

\refitem Walker R.C., Romney J.D., Benson J.M., 1994, ApJ, 430, L45

\refitem Wardle J.F.C., Cawthorne T.V., Roberts D.H., Brown L.F., 1994, 
ApJ, 437, 122

\refitem Whittle M., Saslaw W.C., 1986, ApJ, 310, 104

\refitem Wills B.J., 1996, in Kundt W., ed, Jets from Stars and Galactic
Nuclei. Springer--Verlag, Heidelberg

\refitem Wilson A., 1993, in Burgarella D., Livio M., O'Dea C.P., eds,
Astrophysical Jets. Cambridge Univ. Press, Cambridge, p. 121

\end{document}